\documentclass[twocolumn,prb,aps,showpacs,footinbib,amsmath,amssymb,superscriptaddress]{revtex4-2}
\usepackage{float}
\usepackage{graphicx}
\usepackage{dcolumn}
\usepackage{color}
\usepackage{bm}
\usepackage{natbib,hyperref}
\usepackage{amsmath,amssymb,amsfonts,bbold}
\usepackage[normalem]{ulem}
\usepackage{physics}
\usepackage{listings}
\usepackage[dvipsnames]{xcolor}
\usepackage{color}
\usepackage{braket}

\begin{document}

\title{Unraveling spin entanglement using quantum gates with scanning tunneling microscopy-driven electron spin resonance}

\author{Eric D. Switzer}
\email{eric.switzer@nist.gov}
\affiliation{Donostia International Physics Center (DIPC), 20018 Donostia-San Sebastián, Euskadi, Spain}
\affiliation{Nanoscale Device Characterization Division, National Institute of Standards and Technology, Gaithersburg, Maryland 20899, USA}
\affiliation{Department of Physics, University of Central Florida, Orlando, Florida 32816, USA}
\author{Jose Reina-G{\'{a}}lvez}
\affiliation{Center for Quantum nanoscience, Institute for Basic Science, 03760 Seoul, Republic of Korea }
\affiliation{University of Konstanz,78457 Konstanz, Germany}
\author{G\'eza Giedke}
\affiliation{Donostia International Physics Center (DIPC),  20018 Donostia-San Sebasti\'an, Spain}
\author{Talat S. Rahman}
\affiliation{Department of Physics, University of Central Florida, Orlando, Florida 32816, USA}
\author{Christoph Wolf}
\affiliation{Center for Quantum nanoscience, Institute for Basic Science, 03760 Seoul, Republic of Korea }
\affiliation{Ewha Womans University, 03760 Seoul, Republic of Korea}
\author{Deung-Jang Choi}
\affiliation{Centro de F{\'{i}}sica de Materiales
        CFM/MPC (CSIC-UPV/EHU),  20018 Donostia-San Sebasti\'an, Spain}
\affiliation{Donostia International Physics Center (DIPC),  20018 Donostia-San
 Sebasti\'an, Spain}
 \affiliation{Ikerbasque, Basque Foundation for Science, 48013 Bilbao, Spain}
\author{Nicol{\'a}s Lorente}
\email{nicolas.lorente@ehu.eus}
\affiliation{Centro de F{\'{i}}sica de Materiales
        CFM/MPC (CSIC-UPV/EHU),  20018 Donostia-San Sebasti\'an, Spain}
\affiliation{Donostia International Physics Center (DIPC),  20018 Donostia-San Sebasti\'an, Spain}

\begin{abstract}
Quantum entanglement is a fundamental resource for quantum information processing, and its controlled generation and detection remain key challenges in scalable quantum architectures. 
Here, we numerically demonstrate the deterministic generation of entangled spin states in a solid-state platform by implementing quantum gates via electron spin resonance combined with scanning tunneling microscopy (ESR-STM). 
Using two titanium atoms on a MgO/Ag(100) substrate as a model, we construct a two-qubit system whose dynamics are coherently manipulated through tailored microwave pulse sequences. 
We generate Bell states by implementing a Hadamard gate followed by a controlled-NOT gate, and evaluate its fidelity and concurrence using the quantum-master equation-based code \texttt{TimeESR}. 
Our results demonstrate that ESR-STM can create entangled states with significant fidelity. 
This study paves the way for the realization of atom-based quantum circuits and highlights ESR-STM as a powerful tool for probing and engineering entangled states on surfaces.
\end{abstract}

\date{\today}

\maketitle

\section{Introduction}
Quantum computing relies on the ability to manipulate and entangle quantum states with high fidelity~\cite{Preskill2018,heinrich_quantum-coherent_2021}. 
Among the various platforms proposed for quantum computation, solid-state systems provide a promising avenue due to their scalability and integrability into existing technologies~\cite{Veldhorst2017}. 
One such approach involves using magnetic atoms on insulating substrates, where quantum coherence can be preserved while allowing for controlled quantum operations~\cite{heinrich_quantum-coherent_2021,Yang2019,Wang_2023_Science}. 
In this context, the combination of electron spin resonance with scanning tunneling microscopy (ESR-STM) and atomic manipulation techniques offers a unique method for designing and implementing quantum gates at the atomic scale~\cite{Baumann_Paul_science_2015,Yang2019,Wang_2023,Hong}.

ESR-STM enables the coherent control of individual spins through the application of microwave fields, providing an efficient means to implement quantum logic operations~\cite{Baumann_Paul_science_2015,Yang2019,Seifert_2020}. 
By positioning magnetic atoms on thin insulating layers such as magnesium oxide (MgO) grown on metallic single crystal substrates such as Ag(100), their interactions can be precisely controlled, and their quantum coherence properties can be studied precisely at the level of individual spin states. 
Recent experimental and theoretical advancements have demonstrated that two-qubit quantum gates can be realized by exploiting the interaction between adjacent magnetic adatoms \cite{Yang2019,Wang_2023_Science}. 
Specifically, a controlled-NOT (CNOT) gate, in combination with a Hadamard gate, allows for the deterministic generation of maximally entangled Bell states~\cite{broekhoven_protocol_2024}.

\begin{figure}[!ht]
\includegraphics[width=1.0\linewidth]{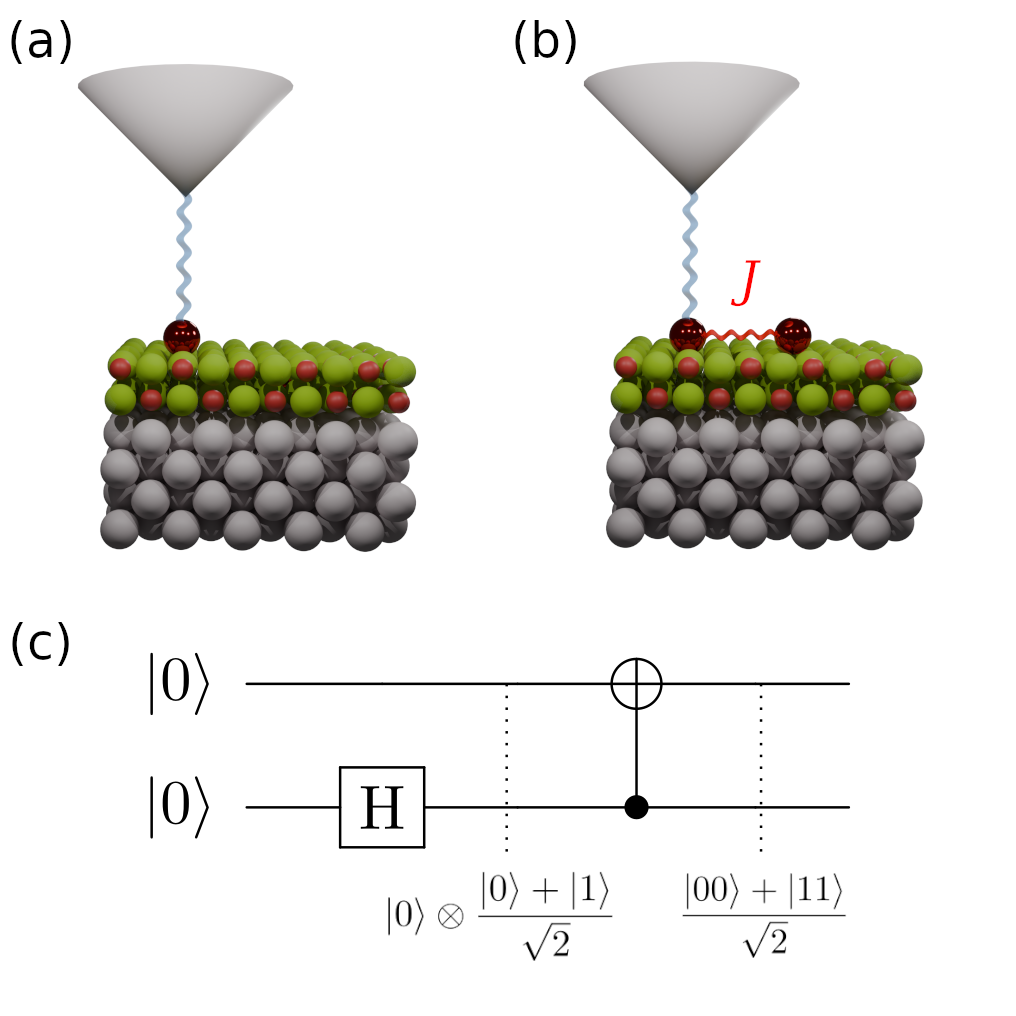}
\centering
\caption{One qubit and two qubit ESR-STM schemes. (a) Atomic scheme fo the ESR-STM setup; one Ti atom ($S=1/2$) on two monolayers of MgO grown on Ag(100). The STM tip is an atomically sharp electrode placed on the Ti atom (designated the transport site), driving the electronic current through it. (b) Scheme of the two-qubit ESR-STM setup consisting of two exchange-coupled Ti atoms (each $S=1/2$) on the same substrate as (a), with the STM tip placed on the transport site. (c) Entanglement gate scheme using a single-qubit Hadamard gate on the second  site, followed by a two-qubit CNOT gate with the second site as the control qubit. The effect of each gate for an input $|0\rangle \otimes |0\rangle$ is shown below the circuit. The final state of the depicted circuit is the Bell state $|\Phi^+\rangle=\frac{1}{\sqrt{2}}(|00\rangle + |11\rangle)$.}
\label{fig1}
\end{figure}

In this work, we numerically demonstrate the realization of a two-qubit quantum gate using ESR-STM to create a Bell state between two titanium (Ti) adatoms located approximately 1.1 nm apart on MgO/Ag(100) (Fig.~\ref{fig1}(b)) and predict realistic time-dependent STM currents using the quantum master equation-derived code \texttt{TimeESR}. 
These atoms have shown to host an effective spin $S=1/2$ orbital. 
An applied external magnetic field splits the $m_{s} = \pm 1/2$ state energies, creating a quantum two-level system.
Our approach utilizes a sequence of pulsed microwave excitations to implement the necessary quantum operations. 
The Hadamard gate is achieved through coherent Rabi oscillations and brings the first qubit into a superposition state. The CNOT gate is implemented by selectively driving a single-spin transition conditioned on the spin state of the control qubits adatom (see Fig.~\ref{fig1}(c)). 
We characterize the performance of the quantum circuit through theoretical simulations and analyze the effects of decoherence due to tunneling currents.
This study demonstrates that ESR-STM can serve as a powerful tool for the implementation of elementary quantum circuits, providing a pathway toward atom-based quantum information processing. 
The ability to create, manipulate, and read out entangled spin states using STM not only advances our understanding of quantum coherence at the atomic scale but also opens up new possibilities for developing quantum technologies on solid surfaces.

\section{Qubit Operations in ESR-STM}
We have created a time-dependent code, \texttt{TimeESR}~\cite{TimeESR}, that models the electron transport and the spin evolution of an arbitrary magnetic atomic or molecular system in contact with two electrodes under realistic out-of-equilibrium conditions, including the effects of electrode voltage, electronic currents and microwave driving. 
Using this code with crafted sequences of tailored microwave pulses, e.g., with the tip of an STM, we can produce controlled spin operations.
In essence, the code allows one to explore how to produce controlled operations on qubits, and the impact of these operations on ESR-STM observables, such as the electronic current. 

In Sections~\ref{sec:TimeESR_model} and \ref{sec:TimeESR_utility}, we provide a brief introduction to the physical model, its implementation in the \texttt{TimeESR} code, and the code's relevant outputs. 
In Sections~\ref{sec:evolution}-\ref{sec:cnot}, we generally outline how one-qubit and two-qubit operations are implemented in ESR-STM. 
Then in Section~\ref{sec:physicalexample}, we show the microwave pulse sequence modeled with \texttt{TimeESR} to entangle two qubits in one of the maximally-entangled Bell states~\cite{Gates}.

\subsection{Model Implemented in the \texttt{TimeESR} Code}
\label{sec:TimeESR_model}
An essential tool employed throughout this work is the numerical code \texttt{TimeESR}, specifically developed to simulate and analyze spin dynamics in ESR-STM experiments. 
This code constitutes the primary computational framework for modeling time-dependent quantum phenomena in single magnetic atoms or molecules placed in an STM junction. 
Its focus lies in investigating the coherent manipulation of localized spin states under the influence of time-periodic driving fields, as mediated by tunneling electrons.

The physical system under consideration consists of a quantum impurity (QI, physically a magnetic adsorbate which may contain one or more magnetic sites) placed in an STM junction, where it is tunnel-coupled to two electronic reservoirs: the metallic tip (electrode $\alpha = \mathrm{T}$) and the substrate ($\alpha = \mathrm{S}$). 
The role of the STM is twofold: it enables charge transport through the adsorbate, and it provides a means to apply time-dependent electric fields that modulate the tunneling rates between the adsorbate and the electrodes. 
By solving the reduced density matrix dynamics under microwave driving and a bias drop between electrodes, \texttt{TimeESR} computes the dynamics of the QI and the evolution of the electronic current that flows through the QI.

The total Hamiltonian of the system is partitioned into three contributions:
    \begin{equation}
    \hat{H}(t) = \hat{H}_{\rm elec} + \hat{H}_{\rm QI} + \hat{H}_{\rm T}(t).
    \label{eq:TotalHamiltonian}
    \end{equation}
The first term, $H_{\rm elec}$, describes the two non-interacting electron reservoirs which model the tip and substrate:
    \begin{equation}
    \hat{H}_{\rm elec} = \sum_{\alpha k \sigma} \varepsilon_{\alpha k} \hat{c}^{\dagger}_{\alpha k \sigma} \hat{c}_{\alpha k \sigma},
    \label{eq:ElectrodeHamiltonian}
    \end{equation}
where $\hat{c}^{\dagger}_{\alpha k \sigma}$ ($\hat{c}_{\alpha k \sigma}$) creates (annihilates) an electron in electrode $\alpha$ with momentum $k$ and spin projection $\sigma \in \{\uparrow, \downarrow\}$ with energy $\varepsilon_{\alpha k}$. 
Each electrode is characterized as a bath with temperature $T_{\alpha}$ and chemical potential $\mu_{\alpha}$. 

The second term, $\hat{H}_{\rm QI}$, is the impurity Hamiltonian and it is given by,
    \begin{equation}
    \hat{H}_{\rm QI} = \sum_{\sigma} \varepsilon_{\sigma} \hat{d}^{\dagger}_{\sigma} \hat{d}_{\sigma}
    + U \hat{n}_{\uparrow} \hat{n}_{\downarrow}
    + \mu_{\rm B} \mathbf{B} \cdot \mathbf{g}\cdot \hat{\mathbf{s}}+\sum_{i=1}^{N}\hat{H}_{\rm S,i},
    \label{eq:AdsorbateHamiltonian}
    \end{equation} 
where $d^{\dagger}_{\sigma}$ ($d_{\sigma}$) creates (annihilates) an electron in a single impurity orbital (henceforth designated as the ``transport'' site) with spin $\sigma$ and energy $\varepsilon_{\sigma}$, and $\hat{n}_{\sigma} = d_{\sigma}^\dagger d_{\sigma}$ is the number operator, $\mu_{\rm B}$ is the Bohr magneton, $\mathbf{B}$ is the local magnetic field on the transport site, $\mathbf{g}$ is the g-tensor, $\hat{\mathbf{s}}$ is the spin operator of the transport orbital, and $\hat{H}_{\rm S,i}$ is the Hamiltonian term that governs all many-body interactions of $N$ magnetic sites not participating in transport.
Equation~(\ref{eq:AdsorbateHamiltonian}) models a many-body entangled system consisting of a single impurity orbital with onsite Coulomb repulsion $U$, under the influence of a local magnetic field $\mathbf{B}$, coupled to additional $N$ magnetic sites.
In our two-site QI model in Fig.~\ref{fig1}(b), Eq.~(\ref{eq:AdsorbateHamiltonian}) corresponds to a transport orbital on the first Ti atom exchange-coupled to the second Ti atom described by $\hat{H}_{\rm S,i}$ with a different local magnetic field.
\texttt{TimeESR} allows for additional complexity of the other magnetic sites $\hat{H}_{\rm S,i}$ in the QI, as explained in the Appendix.

The final contribution, $\hat{H}_{\rm T}(t)$, describes the tunnel coupling between the QI and the electrodes. 
Importantly, this tunneling is modulated by an external time-dependent driving field, which in ESR-STM setups is caused by an  oscillating electric field applied between the tip and the substrate. 
The tunneling Hamiltonian reads,
    \begin{equation}
    \hat{H}_{\rm T}(t) = \sum_{\alpha k \sigma}T_{\alpha}(t) \hat{c}^{\dagger}_{\alpha k \sigma} \hat{d}_{\sigma} + \text{h.c.},
    \label{eq:TunnelingHamiltonian}
    \end{equation}
where $T_{\alpha}(t)$ is the time-dependent, momentum and spin-independent, tunneling amplitude between the impurity and electrode $\alpha$. Following the approach in Refs.~\cite{J_Reina_Galvez_2019, Wolf_C_and_Delgado_F_2020, J_Reina_Galvez_2021}, we keep the time-dependence to its lower order in time,
    \begin{equation}
    T_{\alpha}(t) = T_{\alpha}^{0} \left[ 1 + A_{\alpha} \cos(\omega t + \delta) \right],
    \label{eq:TunnelingModulation}
    \end{equation}
which captures the effect of the modulated tunneling barrier due to the driving field. 
Here, $A_{\alpha}$ is the amplitude of the modulation (assumed small), $\omega$ is the driving frequency, and $\delta$ is a phase shift in the drive. 
Our calculations~\cite{Jose2} have shown that this mechanism is extraordinary efficient in driving the spin, leading to Rabi rates $\Omega$ and coherence times $T_2$ in excellent agreement with experiments.

The full dynamics of the system, including the coupling to the electrodes, is described by a reduced density matrix $\rho_{lj}(t)$, where $l$ and $j$ label the eigenstates of $\hat{H}_{\rm QI}$ by diagonalizing Eq.~(\ref{eq:AdsorbateHamiltonian}). 
We derive the equation of motion for $\rho(t)$ within the Born-Markov approximation~\cite{rammer_2007, Dorn_2021}, treating the coupling $\hat{H}_{\rm T}(t)$ to second order in perturbation theory. 
This results in the time-dependent quantum master equation,
    \begin{align}
    \hbar \dot{\rho}_{lj}(t) &- i \Delta_{lj} \rho_{lj}(t) =  \sum_{vu} \left[ \Gamma_{vl,ju}(t) + \Gamma_{uj,lv}^{*}(t) \right] \rho_{vu}(t) \nonumber \\
    & -  \sum_{vu} \left[ \Gamma_{jv,vu}(t) \rho_{lu}(t) + \Gamma_{lv,vu}^{*}(t) \rho_{uj}(t) \right],
    \label{eq:QME}
    \end{align}
where $\hbar$ is the reduced Planck constant,  $\Delta_{lj} = E_{l} - E_{j}$ is the energy difference between states $l$ and $j$, and $\hbar^{-1}\Gamma_{vl,ju}(t)$ are the time-dependent rates describing tunneling processes involving electron transfer between the impurity and the electrodes.
The real part of $\Gamma_{vl,ju}(t)$ effectively encodes the time-dependent decoherence of the system, while the imaginary part represents the impact of the modulated tunneling on electron transfer.
Further details on the form of the time-dependent rates can be found in Ref.~\cite{J_Reina_Galvez_2021}.

\subsection{Utility of \texttt{TimeESR}}
\label{sec:TimeESR_utility}
The \texttt{TimeESR} code numerically solves Eq.~\eqref{eq:QME} for arbitrary time-dependent tunneling amplitudes and system parameters.
Generally, \texttt{TimeESR} computes the time-dependent current, populations, and spin expectation values. 
As a result, the software captures the essential physics of spin dynamics under time-dependent driving, including tunneling-induced decoherence, non-equilibrium transport, and coherent spin manipulation.

Continuous-wave ESR spectra can also be computed by repeating calculations in \texttt{TimeESR} over a range of different driving frequencies $\omega$, in which each calculation's time propagation is long enough to reach a steady state (dictated by the coherence times of the system studied). 
The DC component of the current, as a function of the driving frequency, is directly comparable with experimental ESR spectra~\cite{Jose2} whilst the time-dependent component is not accessible in the experiment directly due to the slow integrating nature of STM amplifiers, generally limiting the time-resolution to kHz~\cite{Bastiaans2018}. 
The computed current is also accurate at shorter times, permitting the calculation of electronic currents under the presence of short bias pulses. 
Thus, the code is best suited to model time-dependent driving protocols that implement quantum gate operations such as $\pi$-pulses, $\pi/2$-pulses, and more complex sequences designed to achieve universal quantum control, whilst for the long-time limit other methods such as Floquet expansion might be more suitable~\cite{Jose2}.
\texttt{TimeESR} also supports the inclusion of multiple simultaneous or sequential driving frequencies, enabling the study of advanced multi-frequency pulsed protocols used in contemporary ESR-STM experiments. 
These include selective addressing of multiple spins and conditional gate operations akin to two-qubit gates such as the CNOT gate.

\subsection{Single Qubit Unitary Evolution Under ESR Drive}
\label{sec:evolution}
Before addressing the unitary evolution of the two magnetic site quantum impurity in Fig.~\ref{fig1}(b) under ESR-STM, we first describe the time evolution of a simpler problem: a single magnetic site (single qubit), shown in Fig.~\ref{fig1}(a).
We designate the polarized spin ``up'' state $\ket{\uparrow}$ of, e.g., a Ti adatom on MgO/Ag (100)~\cite{Yang2019} as the digital $\ket{0}$, while the ``down'' state $\ket{\downarrow}$ is the digital $\ket{1}$. 
The difference in energy between the two states $\{\ket{0},\;\ket{1}\}$ divided by $\hbar$ is called the Larmor frequency $\omega_{0}/2\pi$.
A harmonic time-dependent interaction will lead to a non-trivial evolution of a generic state. 
In the case of resonance, the driving frequency exactly matches the Larmor frequency, $\omega=\omega_0$. 
Under this condition, the transition probability between the two states oscillates maximally. 
The rate of change of the state, the Rabi rate $\Omega$, measures how fast the transition is driven by the time-dependent interaction.

On resonance in the absence of noise, using $\Omega \ll \omega_{0}$, the time-dependent state can be written as \(|\Psi (t)\rangle = \hat{U} (t) |\Psi (0)\rangle.\) 
Under the above provisos, the unitary in the lab frame can be expressed by the Wigner D-matrix \cite{sakurai_modern_2017} $\bm{D}^{1/2}(\omega_{0}t-\delta',\Omega t,\delta')$ multiplied by an arbitrary phase factor $\exp(i\alpha)$, or equivalently \cite{Orlando},
    \begin{equation}
        \hat{U} (t) 
        \;=\;
        e^{i\alpha}e^{-i\frac{\omega_0}{2}\, t}
        \begin{bmatrix} 
           \cos\!\bigl(\tfrac{\Omega}{2}\,t\bigr) 
           & -\,i\,e^{i\,\delta}\,\sin\!\bigl(\tfrac{\Omega}{2}\,t\bigr) \\[6pt]
           -\,i\,e^{-\,i\,\delta}\,e^{i\,\omega_0\,t}\,\sin\!\bigl(\tfrac{\Omega}{2}\,t\bigr)
           & e^{i\,\omega_0\,t}\,\cos\!\bigl(\tfrac{\Omega}{2}\,t\bigr) 
        \end{bmatrix},
        \label{eq:U_lab_frame}
    \end{equation}
where $\delta = \delta' - \pi/2$,
By transforming to the rotating frame, one obtains the form,
    \begin{equation}
        \hat{U}' (t) 
        \;=\; 
        e^{i\alpha}
        \begin{bmatrix} 
           \cos\!\bigl(\tfrac{\Omega}{2}\,t\bigr) 
           & -\,i\,e^{i\,\delta}\,\sin\!\bigl(\tfrac{\Omega}{2}\,t\bigr) \\[6pt]
           -\,i\,e^{-\,i\,\delta}\,\sin\!\bigl(\tfrac{\Omega}{2}\,t\bigr)
           & \cos\!\bigl(\tfrac{\Omega}{2}\,t\bigr) 
        \end{bmatrix}.
        \label{eq:U_rot_frame}
    \end{equation}
It is in this rotating frame that one can attempt to match the pulsed unitary to a qubit gate operation.
However, Eq.~(\ref{eq:U_rot_frame}) is restrictive in the sense that it alone does not generate all useful qubit gate operations. 
Instead, one must apply one or more pulses of carefully chosen duration $t$, Rabi frequency $\Omega$, and phase $\delta$ to achieve any particular single qubit gate operation. 
If each pulse is ``switched on and off'' sufficiently fast (compared to the timescale $2\pi/\Omega$), the full unitary qubit gate operation may be approximated by a product of discrete pulses,
    \begin{equation}
       \hat{U}_{\text{desired}} \;=\; \hat{U}_n \,\cdots\, \hat{U}_2\,\hat{U}_1,
       \label{eq:PulseSequence}
    \end{equation}
where each $\hat{U}_i$ is of the form Eq.~(\ref{eq:U_rot_frame}) and where each $\Omega_i$ and $\delta_i$ may be distinct.
It is important to note that the typical rise and fall times for ESR-STM pulses are on the order of sub-nanoseconds, which approaches the characteristic timescales of commercially available microwave signal generators operating between 10 GHz and 40 GHz. 
While this does not necessarily invalidate the approximation described above, it does indicate that careful consideration of its limits and conditions is required in this regime.
We next illustrate single- and two-qubit gates via two canonical examples: the Hadamard gate (single-qubit) and the CNOT gate (two-qubit).

\subsection{Single-Qubit Gates with ESR Pulses: Hadamard}
\label{sec:hadamard_example}
The Hadamard gate acts on the time-line of a single qubit and is represented by an ``H'' symbol, Fig.~\ref{fig1}(c). 
The strategy is to design a sequence of pulses given by a product of unitaries, Eq. (\ref{eq:PulseSequence}), such that the final unitary is the Hadamard gate. 
One can achieve this (up to irrelevant global phase) through a sequence of unitaries built from individual ESR-STM pulses that mimic the unitaries of Pauli matrices $\hat{X}$, $\hat{Y}$, $\hat{Z}$ and their fractional powers, e.g., $\text{H} = \hat{Y}^{1/2}\,\hat{Z}$ or $\text{H} = \hat{X}\,\hat{Y}^{1/2}$.
\begin{figure}[!ht]
    \centering
    \includegraphics[width=0.5\linewidth]{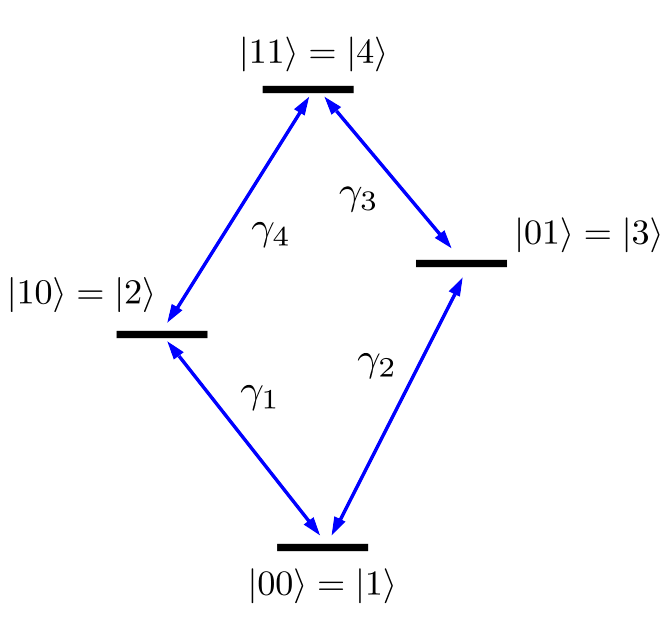}
    \caption{Two-qubit system stemming from two spin-1/2 sites weakly interacting and slightly detuned such that Zeeman-product states, Eq.~(\ref{eqn:Zeeman_states}), are an excellent approximation to the four-level system. The single-qubit transitions between them are designated by their respective rates $\gamma_i$. }
\label{fig2}
\end{figure}
Directly comparing the ESR-STM evolution in Eq.~\eqref{eq:U_rot_frame} to rotations $R(\theta,\phi)=\exp{-\,i\,\tfrac{\theta}{2}\bigl(\cos\phi\,\hat{\sigma}_{x} \,+\,\sin\phi\,\hat{\sigma}_{y}\bigr)}$, one sees that matching phases $\delta$ in ESR pulses effectively generates ``rotations'' around different axes on the Bloch sphere. 
To generate an effective Hadamard gate, one can then follow the procedure,
\begin{enumerate}
  \item A $\pi/2$ pulse ($t = \pi/2\,\Omega$) with phase $\delta = \,-\pi/2$. From Eq.~\eqref{eq:U_rot_frame}, this yields $\hat{U}'_1=\hat{Y}^{1/2}$.
  \item A $\pi$-pulse ($t = \pi/\Omega$) with $\delta = 0$, which is effectively an $\hat{X}$ gate (up to a global phase): $\hat{U}'_2 \approx \hat{X}$.
\end{enumerate}
The combined operation $\hat{U}'=\hat{U}'_2\hat{U}'_1$ results in a Hadamard gate, up to an overall phase factor $i$.
Another route is $\hat{Y}^{-1/2}\,\hat{X}$, achieved by reversing the order and switching the sign of $\delta$. 
Either ESR-STM pulse scheme yields the same final result.

\lstset{
    escapeinside={(*@}{@*)},
}

\begin{figure*}[!ht]
\centering
\begin{lstlisting}
0.000       750.000 ! Initial and final time (ns)
---------------Pulse definition block----------------------------
4                   ! Number of pulses
1                   ! Maximum number of frequencies
(*@\colorbox{red}{\strut\hspace{0.05em}}@*) 0.000     200.000 ! Pulse 1 - times for first pulse (ns)
(*@\colorbox{red}{\strut\hspace{0.05em}}@*) 1.0               ! Pulse 1 - toggle
(*@\colorbox{red}{\strut\hspace{0.05em}}@*) 16.161            ! Pulse 1 - pulse frequency (GHz)
(*@\colorbox{red}{\strut\hspace{0.05em}}@*) 0.0               ! Pulse 1 - phase shift (radians)
(*@\colorbox{blue}{\strut\hspace{0.05em}}@*) 200.000   281.000 ! Pulse 2 - times for second pulse
(*@\colorbox{blue}{\strut\hspace{0.05em}}@*) 1.0               ! Pulse 2 - toggle
(*@\colorbox{blue}{\strut\hspace{0.05em}}@*) 16.161            ! Pulse 2 - pulse frequency
(*@\colorbox{blue}{\strut\hspace{0.05em}}@*) 1.57079633        ! Pulse 2 - phase shift
(*@\colorbox{ForestGreen}{\strut\hspace{0.05em}}@*) 281.000   297.310 ! Pulse 3 - times for the third pulse 
(*@\colorbox{ForestGreen}{\strut\hspace{0.05em}}@*) 1.0               ! Pulse 3 - toggle
(*@\colorbox{ForestGreen}{\strut\hspace{0.05em}}@*) 15.359            ! Pulse 3 - pulse frequency
(*@\colorbox{ForestGreen}{\strut\hspace{0.05em}}@*) 0.0               ! Pulse 3 - phase shift
(*@\colorbox{gray}{\strut\hspace{0.05em}}@*) 297.310   750.000 ! Pulse 4 - times for free evolution
(*@\colorbox{gray}{\strut\hspace{0.05em}}@*) 0.0               ! Pulse 4 - toggle (no driving)
(*@\colorbox{gray}{\strut\hspace{0.05em}}@*) 15.359            ! Pulse 4 - pulse frequency
(*@\colorbox{gray}{\strut\hspace{0.05em}}@*) 0.0               ! Pulse 4 - phase shift
\end{lstlisting}
\caption{Pulse sequence used in the input of TimeESR to produce an entangled $|\Phi^+\rangle$ Bell state. Colors indicate the position of the pulse within the sequence, corresponding with the sequences shown in Fig.~\ref{fig4} and Fig.~\ref{fig5}. }
\label{fig3}
\end{figure*}

\subsection{Two-Qubit Gates with ESR Pulses: CNOT}
\label{sec:cnot}
We now turn to two-qubit operations, focusing on the Controlled-NOT (CNOT) gate, an essential step in creating entangled Bell states. 
Assume a two-qubit system corresponding to the two magnetic sites $T$ (the transport site qubit) and $T'$ (the target site qubit) within the quantum impurity in Fig~\ref{fig1}(b).
In this system, one useful basis set for the Hilbert space described by the states $|T\rangle\otimes|T'\rangle$ is the product states quantized with respect to a particular axis, e.g., aligned to the principal axis of an applied magnetic field: $\{|00\rangle, |10\rangle, |01\rangle,|11\rangle \}$.
Using this basis, the same machinery of Section~\ref{sec:evolution} can then applied to transition between these states.
The effective operation of the CNOT gate with respect to the ``control'' qubit $T$ is to flip the target qubit $T'$ if $\ket{T}=\ket{1}$; otherwise, it does nothing.
In ESR-STM systems with multiple spins, an appropriate pulse frequency $\Omega_{C}$ can selectively drive the transition
\[
   \ket{\sigma_{1}\ldots \downarrow_{T}\ldots \sigma_{T'}\ldots \sigma_{n}}
   \;\longleftrightarrow\;
   \ket{\sigma_{1}\ldots \downarrow_{T}\ldots \bar{\sigma}_{T'}\ldots \sigma_{n}}
\]
only when $\ket{\sigma_{T}}$ is $\ket{\downarrow}$. 
Thus, carefully engineered $\pi$-pulses at $\Omega_{C}$ implement the conditional ``flip'' on the second qubit, yielding a CNOT gate.

\subsection{Physical Example: Generating a Bell State}
\label{sec:physicalexample}
As a concrete illustration, suppose we have a two-spin system with many-body basis states that correspond to product states of the first magnetic site, the transport site, with the second magnetic site with polarization aligned to a Zeeman axis set by locally-applied magnetic fields,
\begin{equation}
\label{eqn:Zeeman_states}
    \{\ket{1}, \ket{2}, \ket{3}, \ket{4}\} = \{\ket{\downarrow \downarrow}, \ket{\uparrow \downarrow}, \ket{\downarrow \uparrow}, \ket{\uparrow \uparrow}\},
\end{equation}
whose spin to digital mapping is $\ket{\downarrow}\rightarrow\ket{0}$ and $\ket{\uparrow}\rightarrow\ket{1}$.
These states can be simulated by using the parameters described in the Appendix, resulting in frequencies (in GHz),
    \begin{align*}
    \omega(1) &\;=\; 0, 
    &\quad \omega(2) &\;\approx\; 15.473,\\
    \omega(3) &\;\approx\; 16.161, 
    & \omega(4) &\;\approx\; 31.520.
    \label{Zeeman}
    \end{align*}
Figure~\ref{fig2} shows a scheme of the energy levels and single-qubit transitions of this two-spin system, assuming the states are sufficiently ``Zeeman-like,'' corresponding with well-known experimentally-accessible systems~\cite{Y_Bae_advanced_science_2018,Wang_2023_Science,Hong}.
The system contains additional configurations describing different transient charge states of the transport site (the unoccupied and doubly-occupied states) in order to account for the electron transport process. 
For the operations on the spins that we describe, we only consider the above four states that correspond to the longer-lived charge state.

\begin{figure}[H]
    \centering 
    \includegraphics[width=.90\linewidth]{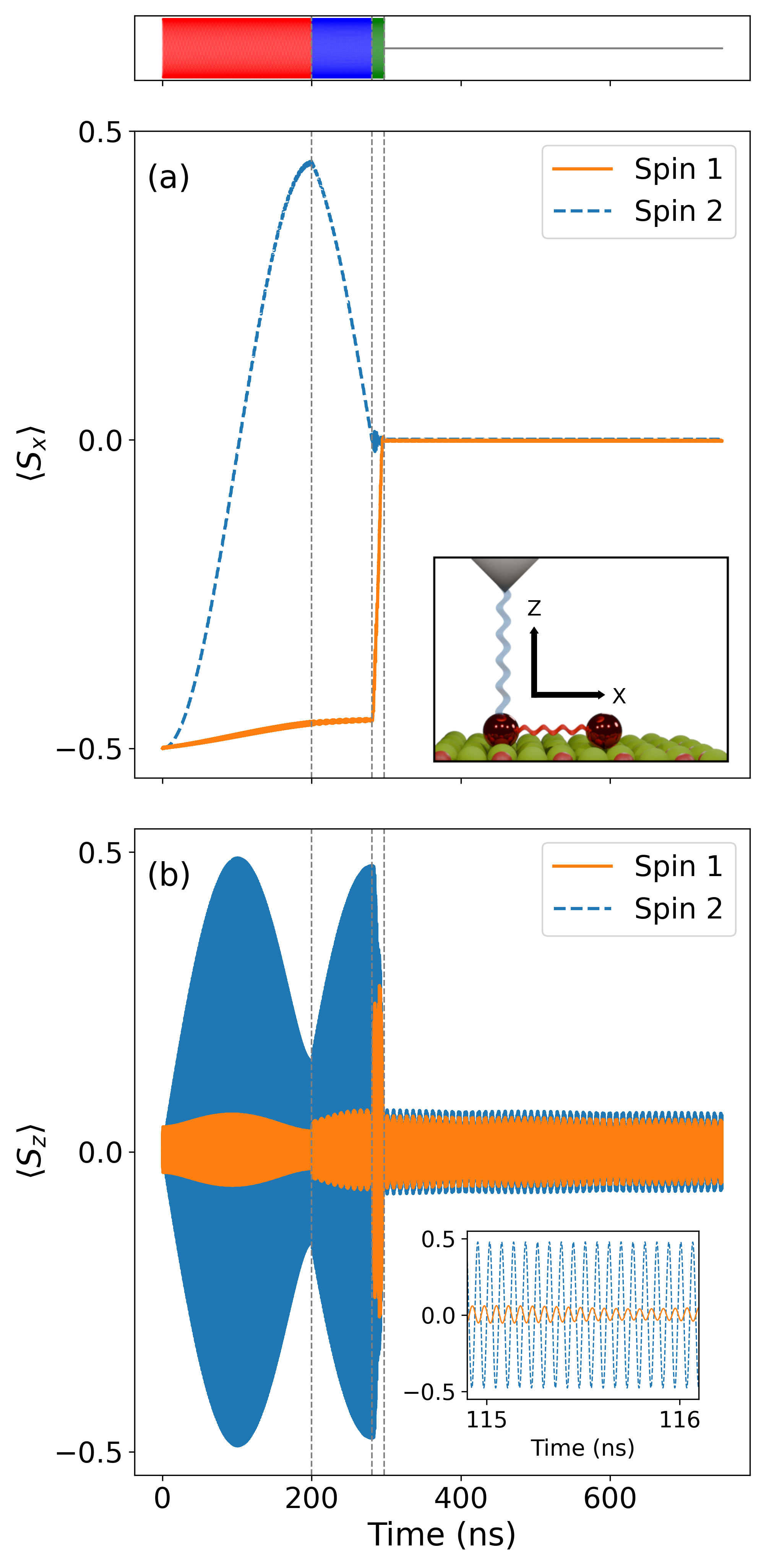}
    \caption{Spin evolution during the quantum circuit execution. The top color bar represents a schematic of the four pulse regions described in Fig.~\ref{fig3}. (a) Expectation value of the spin operator aligned to the locally-applied magnetic field $\hat{S}_{x}$ for each site; see the inset for a description of the principal axes. Initially the frequency is tuned to drive the second site to a superposition state. During this time, no operation is performed on the transport site, but the electronic current causes decoherence and the value of the spin slightly drifts away from $\expval{S^{x}_1}=-0.50$. At 281 ns, the CNOT gate is applied and both expectation values go to zero. (b) Expectation value of the spin operator aligned to the electrode's spin polarization (Z-axis) $\hat{S}_{z}$ for each site.  The expectation value of $\expval{S^{y}}$ (not shown) follows the same pattern as $\expval{S^{z}}$. The profile of $\expval{S^{z}}$ tracks with the result of each pulse operation. As shown in the inset, all calculations are done in the lab frame, leading to oscillations at the Larmor frequency of the in-plane spin expectation values.}
    \label{fig4}
    \end{figure}

Using \texttt{TimeESR}, we identify drive frequencies near the Larmor frequency $\omega_{ij}=\omega(j)-\omega(i)$ near each relevant transition energy, each maximizing population transfer between $\ket{i}$ and $\ket{j}$. 
For example, $\omega_{13}\approx 16.161\,\mathrm{GHz}$ drives $\ket{1}\leftrightarrow \ket{3}$ nearly perfectly (see Fig.~\ref{fig4}(a) and \ref{fig5}(a); at $t\approx200$ ns the spin on the second site flips to almost $+0.50$ and consequently the population largely shifts from state $|1\rangle$ to $|3\rangle$), if the time duration of the pulse corresponds to half a Rabi period. This is called a $\pi$ pulse. 
In Fig.~\ref{fig3} the first pulse is this $\pi$ pulse with duration $t_{\mathrm{pulse}}=\pi/\Omega_{13}\approx 200$~ns, where $\Omega_{13}$ is the Rabi frequency for the oscillation $|1\rangle \leftrightarrow |3\rangle$. 
The Rabi frequencies are determined numerically by plotting the time-dependence of the populations over time at resonant driving.

In this system, a possible sequence to create the Bell state $\ket{\Phi^+} = \tfrac{1}{\sqrt{2}}(\ket{00} + \ket{11})$ is:
\begin{enumerate}
    \item Initialize the system in the ground state $\ket{00}$.
    \item $\hat{X}$ on site 2: apply a $\pi$ pulse at $\omega=\omega_{13}\approx 16.161\,\mathrm{GHz}$ addressing the transition $\ket{1}\leftrightarrow \ket{3}$ with no phase shift. The resulting state is $-i\ket{01}$.
    \item $\hat{Y}^{-1/2}$ on site 2: apply a $\pi/2$ pulse at the same $\omega_{13}$ frequency with a phase shift $\delta=\pi/2$. The result is a state $-\tfrac{i}{\sqrt{2}}(\ket{00}+\ket{01})$.
    \item CNOT with site 2 as the control qubit: apply a $\pi$ pulse for the $\ket{3}\leftrightarrow \ket{4}$ transition at $\omega=\omega_{34}\approx 15.359\,\mathrm{GHz}$ with no phase shift. The resulting state is the Bell state $\ket{\Phi^{+}}$ with a global phase $-i$.
\end{enumerate}
Steps (2) and (3) together implement the Hadamard-like operation on the second site, while step (4) implements a CNOT-like flipping of the transport site, conditional on the second site's state as the logical $\ket{1}$.
Figure~\ref{fig3} is a snapshot of the input of \texttt{TimeESR} needed to implement the above sequence of pulses. 
In the first line the input establishes the total time of the simulation, 750 ns in this example. 
The next line of the input declares the number of pulses, and the maximum number of driving frequencies $\omega_{ij}$ per pulse (in our example, there is only one frequency per pulse).
Next, the four pulses are described by declaring the time interval where it acts, an on/off toggle switch value (1.0 for on, 0.0 for off), its frequency in GHz, and the phase of the pulse in radians. 
Numerical precision of the inputs and a small time-step for the time propagation are important because the decoherence of the spins is fast and quantum operations quickly become noisy.

In ESR-STM, the combination of (i) distinct spin-resonance frequencies $\omega_{ij}$ for different qubit sites and (ii) the ability to introduce phase shifts $\delta$ and fine-tune pulse durations $t$ allows one to implement universal one- and two-qubit operations \cite{Wang_2023_Science,Wang_2023}.
Single-qubit gates such as the Hadamard gate can be constructed from a pair of carefully phased pulses, while two-qubit gates like CNOT can be realized by single-frequency pulses, since every transition frequency is naturally conditional on the control-qubit state (i.e., $\gamma_1$ is different from $\gamma_3$ since they differ by the state of the transport site qubit). 
These building blocks enable the generation of important entangled states, including Bell states.

\section{Creation of Bell states with ESR-STM on two weakly-coupled spins}

Figure~\ref{fig4} shows the spin dynamics over the two sites when the sequence of pulses described in Fig.~\ref{fig3} is performed on the ground state of the spin dimer. 
The  magnetic field is locally applied along the X-axis, which gives the quantization axis of our system. 
The first pulse is a $\pi$-pulse leading to the transition $|00\rangle\rightarrow i|01\rangle$. 
Fig.~\ref{fig4}(a) shows this first pulse, in which the second site's spin expectation value along the quantization axis $\expval{S^{x}_{2}}$ transitions from $-0.50$ to almost $+0.50$.
These simulations show that it is virtually impossible to have a perfect single-qubit $\pi$ pulse in a multi-qubit system due to the complex combined time-evolution of the exchange-coupled spins. 
Our simulations include a junction current which causes decoherence that can be seen in the reduction of the transport site's spin expectation value $\expval{S^{x}_{1}}$ in Fig.~\ref{fig4}(a), from $-0.50$ to approximately $-0.46$ within the time region of the first pulse.

After the second pulse, the $\pi/2$ pulse at the $|00\rangle\rightarrow|01\rangle$ transition frequency, the $\pi$ pulse on the $|01\rangle\rightarrow|11\rangle$ transition is turned on at 281 ns. 
As a consequence, we see that the spin expectation value of the two sites becomes zero along the quantization axis. 
This is an indication that we have created a Bell state, however it is not direct proof as this representation does not directly show the coherent properties of the system. 
To show that we have created Bell states, we detail the fidelity and concurrence of the system in the following section.

\begin{figure}[ht]
    \centering 
    \includegraphics[width=0.9\linewidth]{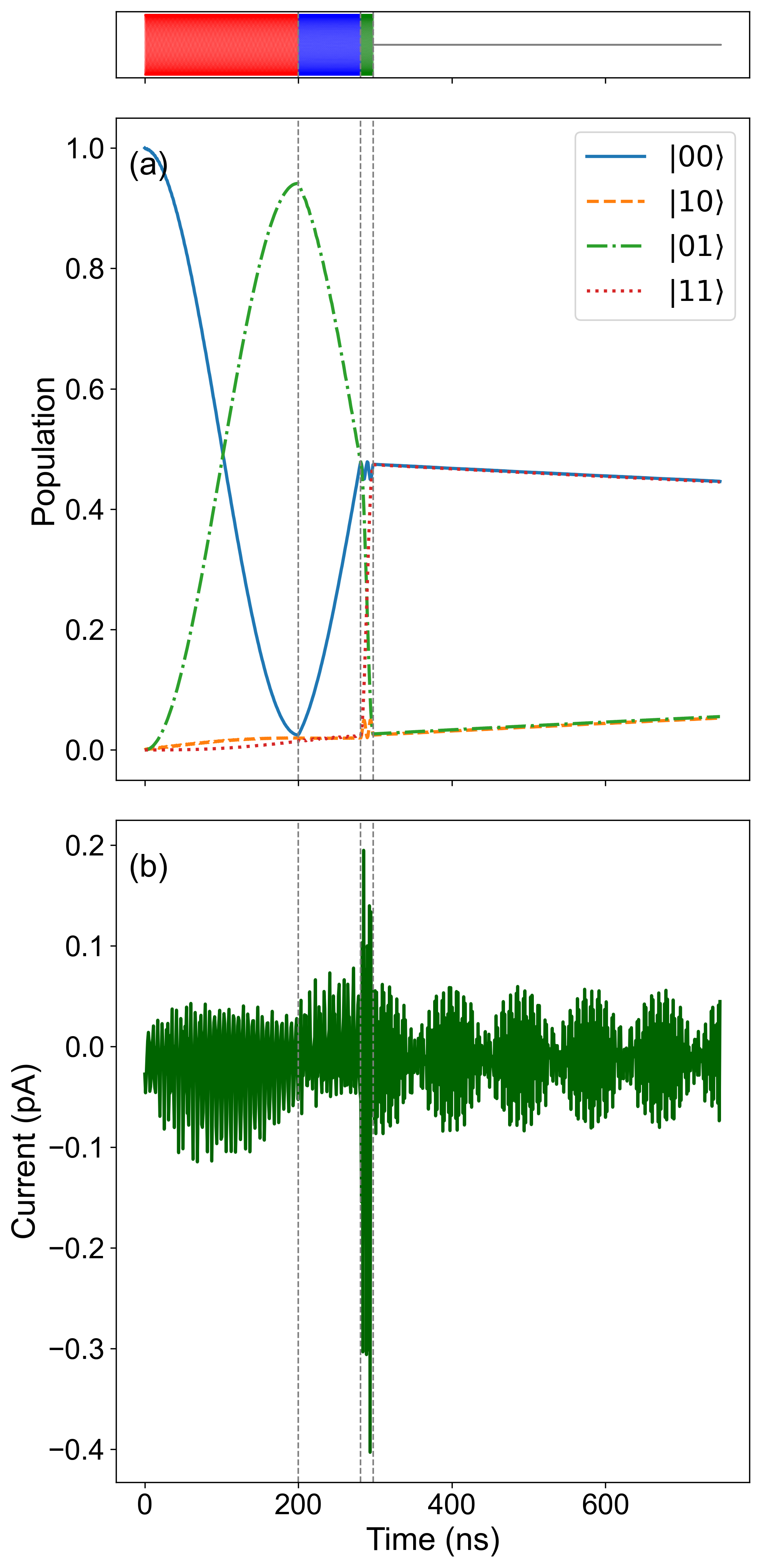}
    \caption{Population of the different states, Fig.~\ref{fig2}, during the quantum circuit execution (a) and the computed electronic current (b). The top color bar represents a schematic of the four pulse regions described in Fig.~\ref{fig3}. Both graphs show the fast evolution taking place before the pulses are turned off at around 300 ns and the free evolution of the two spins is allowed. The population of the states can be easily identified with the expectation value of each single spin in Fig.~\ref{fig4}.}
    \label{fig5}
    \end{figure}

Each spin oscillates in their respective Bloch spheres, as seen in Fig.~\ref{fig4}(b). 
Both the Y and Z components show fast oscillations at the Larmor frequency (see the inset of Fig.~\ref{fig4}(b)), producing complete turns around the X-axis until the final $\pi$ pulse.
After this pulse, the oscillations are greatly reduced in amplitude.
Figure~\ref{fig5}(a) shows the populations of the four states during the realization of the circuit. 
These populations follow the sequence of pulses, and because to the simple form of the quasi-Zeeman states of the quantum impurity Hamiltonian, one can rationalize the values of $\expval{S^{x}_{1}}$ and $\expval{S^{x}_{2}}$ in Fig.~\ref{fig4}(a), based on the population of each state.

Finally, Fig.~\ref{fig5}(b) shows the electronic current that is driven through the transport spin. 
The division of current is apparent before and after the pulse at around 300 ns because it separates the driven and free evolution of the two-spin system.
The current appears noisy but contains clear patterns that reflect the pulses and the dynamic response of the spin system. 
Unfortunately, the time scale of the fluctuations is too fast to allow a direct detection in STM~\cite{Bastiaans2018}. 
Accumulated statistics of the time-averaged current from a large number of consecutive realizations of this gate sequence might allow one to reconstruct the dynamics.

\subsection{Quality of the Bell States}
\begin{figure}[ht]
    \centering 
    \includegraphics[width=0.9\linewidth]{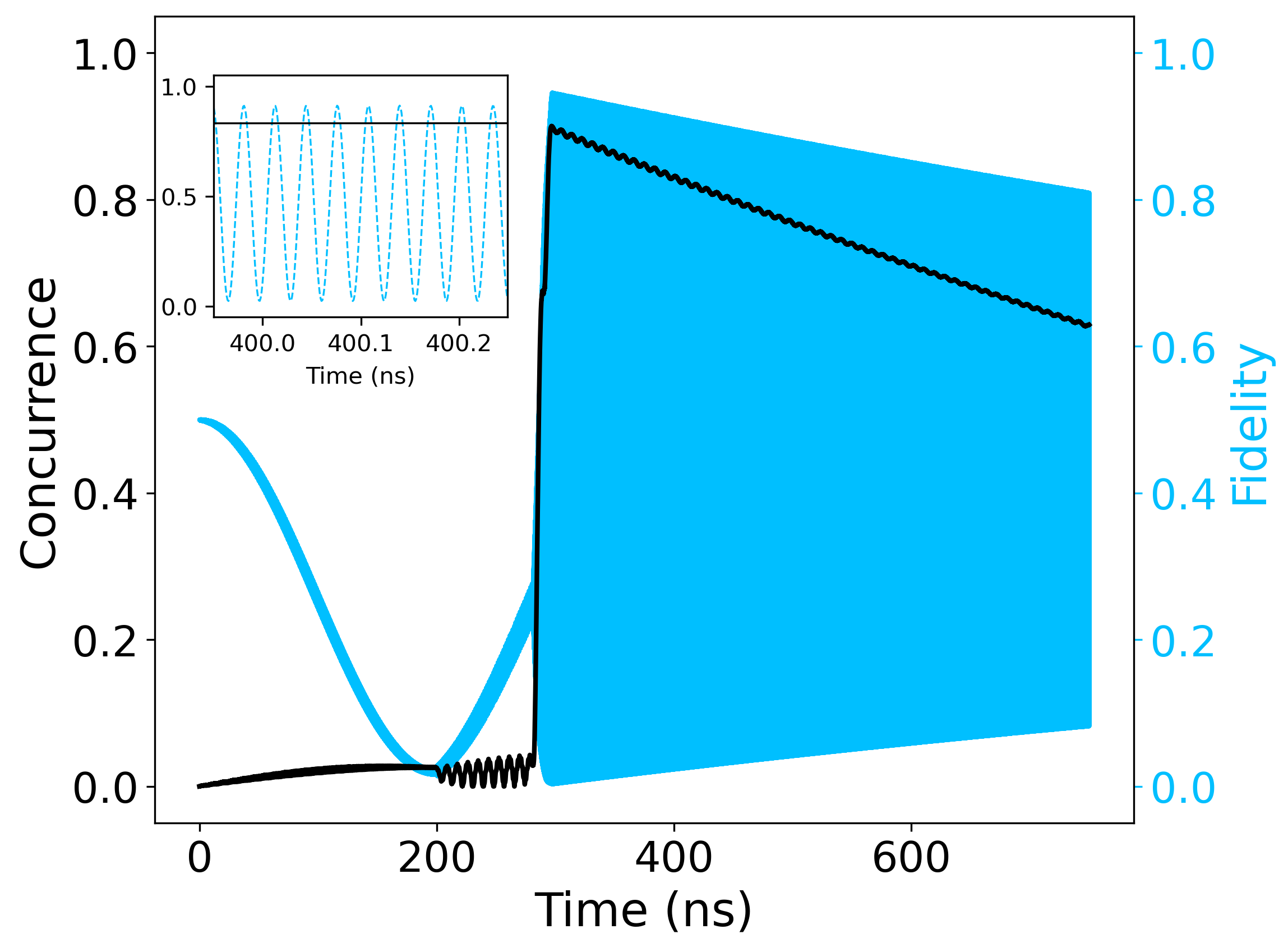}
    \caption{Concurrence and fidelity with respect to $\ket{\Phi^{+}}$ during the execution of the quantum circuit. The concurrence is a measure of the entanglement, accordingly it remains very low until it maximizes at the formation of the $\ket{\Phi^{+}}$. The formation of the latter is monitored through the fidelity, which is the projection of the Bell state on the instantaneous state of the circuit. When the CNOT gate is created, the Bell state is formed and consequently the fidelity reaches 93~\%. As shown in the inset, all calculations are performed in the lab frame which result in fast oscillations at a frequency equal to the difference in the $|00\rangle$ and $|11\rangle$ eigenenergies. Current-induced decoherence is evident in the decrease of concurrence over a scale of $\mu$s after the Bell state is formed.}
    \label{fig6}
    \end{figure}

We quantify the quality with which our circuit prepares the desired Bell state using the fidelity,
    \begin{equation}
    F = \langle\Phi^+|\rho|\Phi^+\rangle,
    \end{equation}
where $\rho$ denotes the state of the two qubits. 
Figure~\ref{fig6} shows the fidelity of our prepared state with the target Bell state $|\Phi^+\rangle$.
We see that at the moment of pulsing the CNOT gate, we create a state that has a fidelity above 90~\%. 
However, the fidelity oscillates rapidly (see the inset of Fig.~\ref{fig6}) at a frequency close to the $31.520$ GHz energy difference between the contributing states to  $\ket{\Phi^+}$ and $\ket{\Phi^-}$, namely $\ket{00}$ and $\ket{11}$.  The fidelity decays over a time scale of $\mu$s as $\rho$ evolves into a mixed state. 
Because the final state is a mixed state, there is always some remnant weight on the Bell state. 

A better insight in the entanglement properties of our system is provided by the concurrence $\mathcal{C}$ \cite{Wootters01}, since it is not affected by the rapidly-oscillating relative phase between the two eigenstates. 
$\mathcal{C}$ takes values between $0$ and $1$. 
For two qubits, $\mathcal{C}=0$ holds for all separable states, while $\mathcal{C}=1$ implies a maximally entangled state (i.e., the state $\ket{\Phi^+}$ up to local unitaries). 
The concurrence gives an upper bound to the possible Bell-state fidelity $(1+\mathcal{C})/2\geq F$ \cite{VeVe02}\footnote{This relationship is via the ``maximal singlet fraction'' $\mathcal{F}=\mathrm{max}_{\psi}\{\bra{\psi}\rho\ket{\psi}\}$, where the maximum is taken over all maximally entangled states. This quantity (for which our $F$ is a lower bound) satisfies $\mathrm{max}\{(1+\mathcal{C})/4,\mathcal{C}\}\leq\mathcal{F}\leq(1+\mathcal{C})/2$ \cite{VeVe02}. Given that $F\leq\mathcal{F}$ the upper bound also holds for $F$. The maximum concurrence $\mathcal{C}=0.900$ and fidelity $F=0.946$ that can be read off from Fig.~\ref{fig6} almost saturate that bound.}. 
Both quantities show that a highly entangled state close to the Bell state was achieved. As $\rho$  evolves, current-induced decoherence accumulates. This results in the decline of the concurrence at a rate larger than the decay of the fidelity envelope. Like the fidelity envelope, the decay occurs over an experimentally-reasonable $\mu$s scale, and is an order of magnitude larger than the slowest gate operation of the circuit.

\section{Conclusions}

We have demonstrated the theoretical realization of universal quantum gate operations in a two-qubit system formed by titanium atoms on a MgO/Ag(100) surface, manipulated using ESR-STM techniques. 
By designing and applying sequences of microwave pulses, we successfully implemented a Hadamard gate and a controlled-NOT (CNOT) gate, which led to the formation of maximally entangled Bell states.
Our numerical simulations, performed with the \texttt{TimeESR} code, capture the time-dependent spin dynamics of the system under realistic experimental conditions. 
We quantified the quality of the entangled states by computing both the fidelity and the concurrence, reaching values above 90~\% before decoherence effects, which happen on a  time scale of $\mu$s, set in. 
The influence of tunnel-induced decoherence was analyzed, demonstrating its impact on the long-term stability of entangled states and the importance of optimizing gate sequences and pulse parameters to mitigate these effects.

By using the numerical results of \texttt{TimeESR} in the time-dependent quantum master equation formalism, the results shown here go beyond prior theoretical studies of entanglement generation in ESR-STM \cite{Castillo,broekhoven_protocol_2024}, and more generalized tripartite spin systems in which the transport site functions as an entanglement witness \cite{switzer_theoretical_2023}.
Specifically we show the crucial impact of the tunneling processes in ESR-STM on system properties and experimental observables during entanglement generation. 
Our results also show use of a transport spin and an exchange coupled second spin are sufficient for quantum gate operations in ESR-STM within the available coherence time of the transport spin.

The work presented here establishes ESR-STM as a viable platform for the implementation of elementary quantum circuits at the atomic scale. 
The precise control of individual spins and their coherent coupling opens promising avenues for developing atomically defined quantum devices. 
Challenges remain, however, in the scalability of the platform (e.g., dynamically tuning the coupling between magnetic sites in the quantum impurity), and the generation of entanglement over a larger number of magnetic sites. 
Future work will focus on extending this approach to larger qubit arrays, and exploring more complex gate sequences. This may require improved coherence times and Rabi rates through optimized surface preparation and quantum control~\cite{reina2024efficientdrivingspinqubitusing}.
Our findings contribute to the growing field of quantum coherence and entanglement in atomic-scale solid-state systems, highlighting ESR-STM as a unique method for the realization of atomic-scale quantum information circuits on surfaces.

\textbf{Acknowledgments.} We thank Eite Tiesinga, Robbie J. G. Elbertse, and Emily A. Townsend for helpful discussions. 
E.D.S. and T.S.R. acknowledge support in part from the Department of Energy, grant number DE-FG02-07ER46354. C.W. acknowledges support by the Institute for Basic Science (IBS-R027-D1).
Any mention of equipment, instruments, software, or materials does not imply recommendation or endorsement by the National Institute of Standards and Technology.
NL thanks projects  PID2021-127917NB-I00 by MCIN/AEI/10.13039/501100011033, IT-1527-22 by Basque Government, ESiM project 101046364 by EU. Views and opinions expressed are however those of the author(s) only and do not necessarily reflect those of the EU. Neither the EU nor the granting authority can be held responsible for them. 
\section*{Appendix: Spin Hamiltonian used in \texttt{TimeESR}}

\texttt{TimeESR} allows for a generalized treatment of the spin Hamiltonian for each magnetic site $\hat{H}_{\text{S},i}$ connected to the transport site in the quantum impurity Hamiltonian $\hat{H}_{\text{QI}}$ of Eq.~(\ref{eq:AdsorbateHamiltonian}).
The general form of the spin Hamiltonian in \texttt{TimeESR} for spin site $i$ with spin $S_{i}$ is,
\begin{eqnarray}
    \label{eqn:gen_spin_hamiltonian}
    \hat{H}_{\text{S},i} &=& \hat{H}_{\text{Z},i} + \hat{H}_{\text{J},i} + \hat{H}_{\text{A},i},
\end{eqnarray}
which accounts for the Zeeman, exchange interaction, and magnetic anisotropy Hamiltonian terms, respectively.
The Zeeman term
$\hat{H}_{\text{Z},i}$ is,
\begin{eqnarray}\label{eqn:hamil_zeeman}
    \hat{H}_{\text{Z},i}&=&\sum_{\chi} \mu_{\rm B} B^{\chi}_{i} g_{i
    \chi} \hat{S}^{\chi}_{i},
\end{eqnarray}
where the sum is over all lab frame directions $\chi$, $B^{\chi}_{i}$ is a locally applied magnetic field, and the g-tensor is simplified by considering a principal-axis dependent g-factor $g_{i\chi}$. 

$\hat{H}_{\text{J},i}$ represents the exchange interaction Hamiltonian between all impurities and impurity site $i$,
\begin{eqnarray}\label{eqn:hamil_exchange}
    \hat{H}_{\text{J},i}&=&\sum_{j\ne i}\sum_{\chi} J^{\chi}_{ij}\hat{S}^{\chi}_{i}\hat{S}^{\chi}_{j},
\end{eqnarray}
where $J^{\chi}_{ij}$ is the exchange interaction strength for the pairing of site $i$ and $j$ in the lab-frame direction $\chi$.
Relevant to all impurities where $S_{i}>1/2$, $\hat{H}_{\text{A},i}$ is a magnetic anisotropy term built from Stevens operators,
\begin{align}\label{eqn:hamil_aniso}
    \nonumber
    \hat{H}_{\text{A},i} &= B^{0}_{2,i}\hat{O}^{0}_{2}(\mathbf{S}_{i})+B^{2}_{2,i}\hat{O}^{2}_{2}(\mathbf{S}_{i})\\
    &\;\;\;\;\;+B^{0}_{4,i}\hat{O}^{0}_{4}(\mathbf{S}_{i})+B^{4}_{4,i}\hat{O}^{4}_{4}(\mathbf{S}_{i}),
\end{align}
where $B^{q}_{k,i}$ is the coefficient of the Stevens operator $\hat{O}^{q}_{k}(\mathbf{S}_{i})$ for impurity site $i$, order $k$, and degree $q$. 

\section*{Appendix: Spin Hamiltonian and \texttt{TimeESR} Hamiltonian Inputs for Bell State Generation}
For the example given in the main text for a single transport site exchange-coupled to another $S=1/2$ impurity, Eq.~(\ref{eqn:gen_spin_hamiltonian}) is simplified considerably.
First, the Zeeman terms are aligned perpendicular to the electrode quantization axis which we take to be the Z-axis, and the $g$ factors are isotropic $g_{1}=g_{2}=g=2$,
\begin{eqnarray}\label{eqn:hamil_zeeman_simplified}
    \hat{H}_{\rm Z}&=&g\mu \left( B^{x}_{1}\hat{S}^{x}_{1} +B^{x}_{2}\hat{S}^{x}_{2}\right).
\end{eqnarray}
The exchange interaction Hamiltonian is simplified to be isotropically ferromagnetic,
\begin{eqnarray}\label{eqn:hamil_exchange_simplified}
    \hat{H}_{\rm J}&=&J\hat{\mathbf{S}}_{1}\cdot\hat{\mathbf{S}}_{2}.
\end{eqnarray}
The magnitude of $B$ and $J$ are set so that the Zeeman term is two orders of magnitude larger than the exchange interaction Hamiltonian, and the occupation energy $\varepsilon$ is negative and two orders of magnitude larger than the Zeeman term.

The values that we have used in the simulations of Fig.~\ref{fig3}, \ref{fig4}, \ref{fig5}, and \ref{fig6} are $\epsilon_d=-5.0$~meV, $U=50$~meV, $B_1^x=0.5509026$~T, $B_2^x=0.5751223$~T, $J=-0.11390$~GHz, and the bias drop is symmetric with magnitude 6.0 mV. 
The rates due to the coupling with the electrodes are 5.0 and 1.0 $\mu$eV with the sample and tip electrodes, respectively.
The drive has been chosen to correspond to 50~\% of the rate with the tip.
The temperature is fixed for both electrodes at $T_\alpha=0.05$ K and 
the tip is 100~\% spin polarized along the Z-axis.

\bibliography{references}
\end{document}